 \documentclass[final,5p,times,twocolumn,authoryear]{elsarticle}

\usepackage{amssymb}
\usepackage{lipsum}
\usepackage{amsmath}
\usepackage{xcolor}

\journal{Physics Letters B}

\begin{document}

\begin{frontmatter}



\title{Probing black holes extremality bounds with non-extensive entropies}


\author[first]{Javier Cancino}
\ead{javier.cancino@alumnos.udg.mx}
\affiliation[first]{organization={Facultad de Física, CUCEI, Universidad de Guadalajara},
            addressline={Blvd. Gral. Marcelino García Barragán 1421, Olímpica}, 
            city={Guadalajara},
            postcode={44430}, 
            state={Jalisco},
            country={México}}

\author[last]{Roberto Santos-Silva}
\ead{roberto.santos@academicos.udg.mx}
\affiliation[last]{organization={Departamento de Ciencias Naturales y Exactas, CU Valles, Universidad de Guadalajara},
            addressline={Carretera Guadalajara-Ameca Km 45.5}, 
            city={Ameca},
            postcode={46600}, 
            state={Jalisco},
            country={México}}

\author[last]{Pablo Carlos López Vázquez}
\ead{pablo.lopez@academicos.udg.mx}

\author[last]{Aldo Martínez-Merino}
\ead{aldo.martinez@academicos.udg.mx}


\begin{abstract}

Due to the importance of studying quantum-gravity-induced corrections to Einstein's theory of gravity, it is of interest to consider non-extensive 
entropies whose potential applications include identifying such corrections. In this Letter, we investigate possible corrections to a universal 
relation associated with the extremality bound of the mass of a black hole by employing entropies defined as functions of the Bekenstein-Hawking entropy.

\end{abstract}





\end{frontmatter}





\section{Introduction}
\label{introduction}

The original laws of black hole mechanics \cite{Bardeen} opened a new paradigm for approaching these objects from a thermodynamic 
perspective. Having a frame to understand, for instance, how a black hole's accretion disk forms and evolves \cite{Ruffini, Blandford}, up to the proposal 
that Einstein's field equations can be derived from the Clausius relation \cite{Jacobson}, is attributable to this new approach to studying gravity and 
its effects.

In \cite{GoonPenco}, using these thermodynamic arguments, the authors derive a universal relation valid for generic black holes that connects 
extremality bounds of mass and entropy, expressed in Equation \eqref{UniversalRel} below.

On the other hand, the quest for a theory of quantum gravity has led to several proposals for obtaining corrections to Einstein's equations. Leading 
candidates such as string theory or loop quantum gravity exist alongside approaches proposing that such corrections can be derived from non-extensive 
entropies (see, for example, \cite{Sheykhi1, Sheykhi2, Obregon}). It is worth noting previous work conducted regarding the use of non-extensive entropies 
like \cite{MM1, MM2, MM3}.

Interestingly, from this perspective, Eq. \eqref{UniversalRel} can be tested to investigate whether any violation of this relation occurs when these 
non-extensive entropies are considered. In general, such entropies depend on a characteristic parameter; thus, our aim in this letter is to examine how 
this universal relation behaves under non-extensive entropies, as well as the variation of extremal quantities when this characteristic parameter varies. 
We note that \cite{Anand} address this same scenario for the example presented in the original reference of \cite{GoonPenco}; we will comment on their 
results in the Discussion section.

Because part of our interest lies in considering corrections originating from quantum gravity, after reviewing the universal relation of \cite{GoonPenco}, 
we proceed by examining entropy with logarithmic quantum gravity corrections, subsequently moving to the study of non-extensive entropies. We conclude with a 
discussion of the results obtained here along with those of \cite{Anand}.


\section{A universal relation for extremality bounds of black holes}

Using thermodynamic arguments, in the usual theory of General Relativity, \cite{GoonPenco} derived a universal relation between the extremality bound 
of the mass of a black hole and the variation of its entropy at the same value,
\begin{equation}\label{UniversalRel}
    \frac{\partial M_{ext} (\vec{Q}, \epsilon)}{\partial \epsilon} = \lim_{M \rightarrow M_{ext}( \vec{Q}, \epsilon)} -T \left( 
    \frac{\partial S (M, \vec{Q}, \epsilon)}{\partial \epsilon}\right)_{M, \vec{Q}}, 
\end{equation}
where $M_{ext}$ is this extremality bound and $\vec{Q}$ represents all the parameters on which the black hole depends;
$\epsilon$ is the parameter that we are varying and $S$ is the entropy of the black hole.

This universal relation was tested on the charged black hole in $AdS_4$ spacetime, whose line element, from the classical Euclidian action, is ($c = G = 1$)
\begin{subequations}
\begin{align}
    ds^2 &= \Sigma(r) dt^2 + \Sigma(r)^{-1} dr^2 + r^2 d\Omega_2^2, \\
    \Sigma(r) &= 1 - \frac{2M}{r} + \frac{Q}{r^2} + \frac{r^2}{l^2}, \\
    A_\mu dx^\mu &= \frac{i Q}{r} dt,
\end{align}
\end{subequations}
where $d\Omega_2^2$ is the usual line element of the sphere. For this case, the value of the mass of the black hole is obtained by solving the
equation $\Sigma (r_+) = 0$, which is given by
\begin{equation}
    M_0 = \frac{S_{BH} + \pi Q^2}{2 \sqrt{\pi S_{BH}}} + \frac{S_{BH}^{3/2}}{2 \pi^{3/2} l^2},
\end{equation}
where we use the Bekenstein-Hawking (BH) entropy, $S_{BH} = A/ 4\ell_P$ and $A = 4\pi r_+^2$; hence, rescaling the radius of the 
event horizon to $r_+ \rightarrow r_+ / \ell_P$ such that\footnote{We are using a dimensionless expression for the entropy, $S \rightarrow S/k_B$.}
\begin{equation}
S_{BH} = \pi r_+^2.
\end{equation}
The temperature of the black hole is equal to
\begin{equation}
    T_0 = \frac{\partial M_0}{\partial S_{BH}} = \frac{S_{BH} - \pi Q^2}{4 \sqrt{\pi} S_{BH}^{3/2}} + \frac{3 S_{BH}^{1/2}}{4 \pi^{3/2} l^2}.
\end{equation}
The extreme value for the entropy is obtained by finding the value that causes the temperature to vanish. In this case, the value is
\begin{equation}\label{EntropiaExtrema}
    S_{0, ext} = \frac{\pi l^2}{6} \left( -1 + \sqrt{1 + \frac{12 Q^2}{l^2}} \right).
\end{equation}

As an example, it is shown that the relation \eqref{UniversalRel} is valid when we deform, for instance, the $AdS$ radius $l \rightarrow 
l/ \sqrt{\epsilon}$, for small $\epsilon$. Substituting $S_{0, ext}$ in the expression of the mass, we obtain the extremality bound $M_{ext}$. The 
variation of the resulting quantity wrt $\epsilon$ compared to $-T (\partial S / \partial \epsilon)_{M, Q, l} |_{M_{ext}}$ gives the desired equality. 
In the next section, we probe Eq. \eqref{UniversalRel} for the entropy with the logarithmic correction of quantum gravity.


\section{The quantum gravity entropy}

We perform the calculation with the logarithm correction to the BH entropy by quantum gravity, which is \citep{Fursaev}
\begin{equation}\label{Entropia-cuantica}
    S = S_{BH} - \beta \ln S_{BH} + \gamma S_{BH}^{-1},
\end{equation}
where $\beta$ and $\gamma$ are two small parameters. For now, we make the variation with respect to the $\beta$ parameter taking $\gamma = 0$. These corrections
have been obtained in models of quantum gravity, for instance, String Theory or Loop Quantum Gravity. This is why it is interesting to consider whether the 
universal relation is still valid for this and other non-extensive entropies.

In order to perform the following calculations, we have to express the mass in terms of the entropy. Hence, using the Lagrange inverse theorem \citep{Grossman}
we have
\begin{equation}
    \pi r_+^2 = S + \beta \ln S + \beta^2 \frac{\ln S}{S}.
\end{equation}
The black hole mass is written as
\begin{equation}\label{masa-area}
    M = \frac{\pi r_+^2 + \pi Q^2}{2 \sqrt{\pi} (\pi r_+^2)^{1/2}} + \frac{(\pi r_+^2)^{3/2}}{2 \pi^{3/2} l^2}.
\end{equation}
%


\subsection{Calculations up to first order}

Considering the area of the black hole as a function of entropy, using a Taylor expansion, the mass has the following functional form up to first order
\begin{equation}\label{Qmass}
    M( S ) = M_0 + \beta \frac{\partial M_0}{\partial S} \ln S.
\end{equation}

We vary the expression of the mass wrt entropy in order to find the temperature. In this case, we have that
\begin{equation}
    T = \frac{\partial M}{\partial S} = T_0 + \beta \frac{\partial}{\partial S} (T_0 \ln S),
\end{equation}
and from this we calculate the maximum value of the entropy making $T=0$. Multiplying this expression by $4/3 \times \pi^{3/2} l^2 S^{3/2}$ considering 
vanishing temperature, at first order, we have 
\begin{equation}
    S_{ext} = S_{0, ext} - \beta \frac{4 \sqrt{\pi} S_{ext}^{3/2}}{\sqrt{1 + \frac{12 Q^2}{l^2}}} \left[ \frac{T_0 (S_{ext})}{S_{ext}} + \ln S_{ext} 
    \frac{\partial T_0}{\partial S} (S_{ext}) \right].
\end{equation}
Invoking again the Lagrange inverse theorem, using $T_0 (S_{0,ext}) = 0$, the extreme value of the entropy is
\begin{equation}
    S_{ext} = S_{0, ext} - \beta \frac{4 \sqrt{\pi} S_{0, ext}^{3/2}}{\sqrt{1  + \frac{12 Q^2}{l^2}}} \ln S_{0, ext} \frac{\partial T_0}{\partial S} 
    (S_{0, ext}).
\end{equation}
An explicit calculation gives that
\begin{equation}\label{MiraTu}
    \frac{4 \sqrt{\pi} S_{0, ext}^{3/2}}{\sqrt{1  + \frac{12 Q^2}{l^2}}} \frac{\partial T_0}{\partial S} (S_{0, ext}) = 1.
\end{equation}
Therefore, the extreme value for the entropy is given by
\begin{equation}\label{QExtremeValue}
    S_{ext} = S_{0, ext} - \beta \ln S_{0, ext}.
\end{equation}

When we substitute this value in the mass, up to first order, we see that there is no modification of the extremality bound for this quantity. This is 
$M_{ext} = M_0 (S_{0, ext})$. This result immediately implies that $( \partial M_{ext} / \partial \beta) =0$.

Returning to Eq. \eqref{Qmass} and deriving it wrt $\beta$, after equating the result to zero, we find that
\begin{equation}\label{Qrhs}
    -T \left( \frac{\partial S}{\partial \beta} \right)_{M, Q, l} = T_0 (S) \ln S.
\end{equation}
Observe that if the mass has corrections up to first order in $\beta$, $\partial M / \partial \beta$ is an expression up to zero order. Thus, Eq. \eqref{Qrhs}
has to be calculated up to zero order as well. Therefore, substituting Equation \eqref{QExtremeValue} this expression vanishes. This means that the universal 
relation \eqref{UniversalRel} still holds up to this order.


\subsection{Calculations up to second order}

Since we note that the corrections to the mass, up to first order, give a relatively trivial result for the universal relation, we explore whether, up to the 
next leading order, such relation still holds. The expression \eqref{Qmass} to second order is
\begin{equation}\label{QQmass}
    M( S ) = M_0 + \frac{\partial M_0}{\partial S} \left[ \beta \ln S + \beta^2 \frac{\ln S}{S} \right] + \frac{1}{2} \frac{\partial^2 M_0}{\partial S^2} 
    \beta^2 (\ln S)^2.
\end{equation}
Performing the derivative of this expression wrt $\beta$ we obtain the relation
\begin{equation}
    -T \left( \frac{\partial S}{\partial \beta} \right)_{M, Q, l} = T_0 \ln S + \beta \frac{\partial}{\partial S} (T_0 (\ln S)^2),
\end{equation}
where the corrected temperature is
\begin{equation}
    T = \frac{\partial M}{\partial S} = T_0 + \beta \frac{\partial}{\partial S} (T_0 \ln S),
\end{equation}
as was obtained before. Once more, we note that if the mass has corrections up to second order in $\beta$, then $\partial M/ \partial \beta$ has 
corrections up to first order.

Taking into account the extreme value for the entropy already obtained (Eq. \eqref{QExtremeValue}), the corresponding extremality bound for the mass is
\begin{equation}
    M_{ext} = M_{0,ext} + \frac{1}{2} \beta^2 \frac{\partial T_0}{\partial S} (\ln S_{0, ext})^2 \left( 1 - \frac{4 \sqrt{\pi} S_{0, ext}^{3/2}}{\sqrt{1 + 
    \frac{12 Q^2}{l^2}}} \frac{\partial T_0}{\partial S} \right)^2,
\end{equation}
where the derivatives are evaluated at $S_{0, ext}$. This implies that there is no correction for the extremality value of the mass taking into account
\eqref{MiraTu}. Thus, we again have $\partial M_{ext} / \partial \beta = 0$.

After evaluating the extremal value of the entropy, for the lhs of Eq. \eqref{UniversalRel}, we have the following
\begin{equation}
    -T \left( \frac{\partial S}{\partial \beta} \right) \Big|_{M_{ext}} = \beta \frac{\partial T_0}{\partial S} (\ln S_{0, ext})^2
    \left( 1 - \frac{4 \sqrt{\pi} S_{0, ext}^{3/2}}{\sqrt{1  + \frac{12 Q^2}{l^2}}} \frac{\partial T_0}{\partial S} \right) = 0.
\end{equation}
Therefore, the universal relation \eqref{UniversalRel} still holds.


\section{The case for non-extensive entropies}

In this section, we study the case for the diverse non-extensive entropies we find in the literature. We are interested in finding out if for all these 
expressions the universal relation \eqref{UniversalRel} still holds.


\subsection{Barrow entropy}

This entropy is a power law of the Bekenstein-Hawking entropy with the expression \citep{Barrow}
\begin{equation}
    S_B = (S_{BH})^{1 + \frac{\Delta}{2}}.
\end{equation}
In the rest of the paper, we use the equality $S_{BH} = \pi r_+^2$. Define $\delta^{-1}= 1 + \frac{\Delta}{2}$, and the inverse relation between entropies
is now
\begin{equation}
    \pi r_+^2 = S_B^\delta.
\end{equation}
Substituting in eq. \eqref{masa-area} the expression of the black hole mass is
\begin{equation}
    M (S_B; \delta) = \frac{S_B^{\delta} + \pi Q^2}{2 \sqrt{\pi} S_B^{\delta/2}} + \frac{S_B^{3 \delta/ 2}}{2 \pi^{3/2} l^2}.
\end{equation}
From this expression, we can calculate the temperature, which is
\begin{equation}
    T (S_B; \delta) = \delta \left[ \frac{S_B^{\delta} - \pi Q^2}{4 \sqrt{\pi} S_B^{\delta/2 + 1}} + \frac{3 S_B^{3 \delta/ 2 - 1}}{4 \pi^{3/2} l^2} \right]
    = \delta T_0 (S_B^\delta).
\end{equation}
The value of entropy for which the temperature vanishes is
\begin{equation}
    S_{B, ext}^\delta = S_{0, ext}.
\end{equation}
This implies that the mass of the black hole at its extreme value coincides with the extreme value for the usual entropy, which is $M(S_{B, ext}) = M _{0, ext}$.
Therefore, the mass does not depend on the parameter $\delta$. Hence, trivially, the lhs of Eq. \eqref{UniversalRel} vanishes.

For the rhs of Eq. \eqref{UniversalRel} we have for this entropy that
\begin{equation}
    -T(S, \delta) \left( \frac{\partial S}{\partial \delta} \right)_{M, Q} = S_B^\delta \ln S_B \frac{T_0 (S_B^\delta)}{\delta},
\end{equation}
which also vanishes when the extreme value of entropy is substituted. Therefore, the universal relation \eqref{UniversalRel} holds for this entropy.

The case for the Tsallis-Cirto entropy \citep{Tsallis-Cirto}, $S_{TC} = S_{BH}^{\tilde{\delta}}$, is similar, and the universal relation holds also for this entropy.


\subsection{Rényi entropy}

The expression for this entropy is \citep{Renyi}
\begin{equation}
    S_R = \frac{1}{\lambda} \ln (1 + \lambda S_{BH}),
\end{equation}
and the inverse relation is readily given by
\begin{equation}\label{RenyiEntropy}
    S_{BH} = \pi r_+^2 = h(S_R) = \frac{1}{\lambda} (e^{\lambda S_R} - 1).
\end{equation}
When we substitute this expression in \eqref{masa-area}, and by performing the derivative wrt $S_R$ we find that the temperature is
\begin{equation}
    T_R = e^{\lambda S_R} \left\{ \frac{ \frac{1}{\lambda} (e^{\lambda S_R} - 1) - \pi Q^2}{4 \sqrt{\pi} (\frac{1}{\lambda} (e^{\lambda S_R} - 1))^{3/2}}
    + \frac{3 (\frac{1}{\lambda} (e^{\lambda S_R} - 1))^{1/2}}{4 \pi^{3/2} l^2} \right\} .
\end{equation}
Looking for the value of the entropy that causes the temperature to vanish, we find this value to be
\begin{equation}\label{ExtremaRenyi}
    S_{R,ext} = \frac{1}{\lambda} \ln (1 + \lambda S_{0,ext}),
\end{equation}
where $S_{0, ext}$ is the same as in Eq. \eqref{EntropiaExtrema}.

In order to investigate whether the universal relation \eqref{UniversalRel} still holds for this entropy, we substitute $S_{R, ext}$ into Eq. \eqref{masa-area} 
when $h(S_{R, ext})$. What we arrive at is that the extremality bound of the mass does not depend on the parameter $\lambda$, in fact, we can check that
\begin{equation}
    M(S_{R, ext}) = M_0 (S_{0, ext}).
\end{equation}
Therefore, the lhs of Eq. \eqref{UniversalRel} vanishes.

On the other hand, we can verify that
\begin{equation}
    -T \left( \frac{\partial S}{\partial \lambda} \right)_{M,Q,l} = \left\{ -\frac{1}{\lambda^2} (e^{\lambda S_R} - 1) + \frac{S_R}{\lambda} e^{\lambda S_R} \right\}
    T_0 (f(S_R)).
\end{equation}
Evaluating the expression in the extreme value of entropy \eqref{ExtremaRenyi}, we can easily see that the expression vanishes. Thus, the rhs of Eq. \eqref{UniversalRel}
also vanishes, and the universal relation still holds trivially for this entropy.


\subsection{Non-extensive entropies of the form $S = f(S_{BH}; \epsilon)$}

After studying the two previous cases, the Barrow and Rényi entropies, besides the other two interesting non-extensive entropies Kaniadakis and Sharma-Mittal,
whose principal results are written in \ref{AppA}, we can make a general analysis of what is happening for an entropy of general form 
$S= f (S_{BH}; \epsilon)$ where $\epsilon$ is the parameter which characterizes the given entropy, and why the universal relation \eqref{UniversalRel} holds 
for these classes of entropies, for this particular parameter.

Let us assume that we can invert the relation between the non-extensive entropy and the entropy $S_{BH}$, and express $S_{BH} = \pi r_+^2 = f^{-1} (S; \epsilon)$ 
(all cases treated here satisfy this condition). We have an expression for the mass of the black hole in terms of the entropy when we substitute this inverse 
relation in Eq. \eqref{masa-area},
\begin{equation}\label{masa-no-extensiva}
    M(f^{-1} (S) ; Q, l).
\end{equation}
The temperature of the black hole in this case is given by
\begin{equation}
    T(S; \epsilon, Q,l) = \frac{\partial M( f^{-1}(S); \epsilon, Q, l)}{\partial S} = \frac{d f^{-1}}{d S} T_0 (f^{-1}(S)),
\end{equation}
due to $T_0 = \partial M_0/ \partial S_{BH}$. Since we have $d f^{-1} / d S \neq 0$, looking for the value of the entropy that causes the temperature to 
vanish, $S_{ext}$, we must solve $T_0 (f^{-1}(S_{ext})) = 0$. This has a solution $S_{ext} = f (S_{0, ext})$, as we can attest from the examples. We must 
consider $T_0 (S_{0, ext}) = 0$.

Using the extreme value $S_{ext}$ and substituting it in the expression for the mass of the black hole, we observe that there is no modification of the extreme
value of the mass or equivalently that the mass does not depend on the parameter $\epsilon$, since $M_{ext}(f^{-1} (S_{ext})) = M_0 (S_{0, ext}) = M_{0, ext}$. 
Hence, we have
\begin{equation}
    \frac{\partial M_{ext}}{\partial \epsilon} = 0.
\end{equation}

On the other hand, from Eq. \eqref{masa-no-extensiva}, performing a derivative wrt $\epsilon$, we arrive at the expression
\begin{equation}
    -T \left( \frac{\partial S}{\partial \epsilon} \right)_{M, Q, l} = F(S; \epsilon) T_0(f^{-1} (S)),
\end{equation}
with the function $F$ has no relevance in the present discussion for now. We make a haste to observe that in the extreme value this expression vanishes due to the 
presence of the function $T_0$ evaluated in $S_{ext}$. And with it, the universal relation \eqref{UniversalRel} is fulfilled when the parameter we consider is the 
one that characterizes the entropies.


\section{Discussion}

We acknowledge the relevance of expressions like Eq. \eqref{UniversalRel} when corrections due to quantum gravity effects are discussed, since by this kind of
results one can look for deviations to semi-classical expressions. It is in this spirit that we study the possible corrections due to non-extensive entropies.
We find that there are no corrections when the parameter on which these entropies depend is varied, as was the case also for the entropy which has 
logarithmic corrections due to quantum gravity.

Before we close this section, we want to point out that we can attest that this universal relation is satisfied for the parameter on which other non-extensive 
entropies depend, as well as for the parameter varied in the original paper of \cite{GoonPenco}, which was the cosmological constant or, equivalently, the 
curvature of the $AdS$ spacetime, $l \rightarrow l/\sqrt{\epsilon}$, when the non-extensive entropies are considered. These latter calculations are performed 
explicitly in \ref{AppB} for the Rényi entropy. In contrast to what is stated in reference \cite{Anand}. The authors argue that Eq. \eqref{UniversalRel} is 
not satisfied when non-extensive entropies are considered and the parameter $l$ is varied to $l/\sqrt{\epsilon}$. We noted that they did not consider the 
substitution of the extreme value of the entropy at the expression for the mass of the black hole, for instance. This omission led them to the reported 
negative result.


\section{Summary and conclusions}

In this letter, we analyze the possibility of any deviation of the universal relation \eqref{UniversalRel} when non-extensive entropies are considered and
parameters such as the cosmological constant and those characterizing such entropies are varied. We found that the above mentioned universal relation holds in
all these cases, in contrast to was stated in \cite{Anand} for the non-extensive entropies in particular.


\section*{Acknowledgements}

All authors thank Universidad de Guadalajara for their support through the PROSNII program, as well as SECIHTI for their corresponding support for this 
investigation.

\appendix

\section{Results on Kaniadakis and Sharma-Mittal entropies}
\label{AppA}

In this appendix, we write the expressions for the mass and temperatures for the other two entropies we consider in this investigation.

We have the Kaniadakis entropy \citep{Kaniadakis}, defined by
\begin{equation}
    S_\kappa = \frac{1}{\kappa} \sinh \kappa S_{BH},
\end{equation}
with inverse relation
\begin{equation}
    \pi r_+^2 = \frac{1}{\kappa} \sinh^{-1} \kappa S_\kappa.
\end{equation}

We have for the mass $M (\pi r_+^2 (S_\kappa))$, and for the corresponding temperature, it is
\begin{equation}
    T (S_\kappa) = \frac{1}{\sqrt{1 + \kappa^2 S_\kappa^2}} T_0 ( \pi r_+^2(S_\kappa)).
\end{equation}

Finally, for the Sharma-Mittal entropy \citep{Sharma-Mittal}, whose expression is
\begin{equation}
    S_{ShM} = \frac{1}{\rho} ( [ 1 + \lambda S_{BH}]^{\frac{\rho}{\lambda}} - 1 ),
\end{equation}
which depends on two parameters $\rho$ and $\lambda$. The inverse relation is
\begin{equation}
    \pi r_+^2 = \frac{1}{\lambda} ( [1 + \rho S_{ShM}]^{\frac{\lambda}{\rho}} - 1).
\end{equation}

The expressions for the corresponding mass and temperature are $M_0 (\pi r_+^2 (S_{ShM}))$ and
\begin{equation}
    T (S_{ShM}) = (1 + \rho S_{ShM})^{\frac{\lambda}{\rho} - 1}\; T_0 (\pi r_+^2 (S_{ShM}))
\end{equation}

\section{An explicit calculation with Renyi entropy}
\label{AppB}

As was said in Section 2, we show, for the example of Rényi entropy, that Eq. \eqref{UniversalRel} is valid using this entropy and when the parameter of 
the curvature of the $AdS$ spacetime is varied, $l \rightarrow l/\sqrt{\epsilon}$, for small $\epsilon$.

With the varied parameter, Equation \eqref{masa-area} is
\begin{equation}\label{Rmasa-area}
    M = \frac{\pi r_+^2 + \pi Q^2}{2 \sqrt{\pi} (\pi r_+^2)^{1/2}} + \frac{ \epsilon (\pi r_+^2)^{3/2}}{2 \pi^{3/2} l^2}.
\end{equation}
We calculate the temperature corresponding to this mass, and it is
\begin{align}
    T &= \frac{\partial M(S_R; \epsilon)}{\partial S_R} = \frac{\partial (\pi r_+^2)}{\partial S_R} \frac{\partial M}{\partial (\pi r_+^2)} \nonumber \\
    &= e^{\lambda S_R} \left\{ \frac{\pi r_+^2 - \pi Q^2}{4 \sqrt{\pi} ( \pi r_+^2 )^{3/2}} + \frac{3 \epsilon ( \pi r_+^2 )^{1/2}}{4 \pi^{3/2} l^2 } \right\},
\end{align}
where the area of the black hole is given by the relation \eqref{RenyiEntropy}.

Again, from Eq. \eqref{Rmasa-area}, we can derive the expression
\begin{equation}\label{Rrhs}
    -T \left( \frac{\partial S_R}{\partial \epsilon} \right)_{M, Q, l} = \frac{( \pi r_+^2 )^{3/2}}{2 \pi^{3/2} l^2}.
\end{equation}

In establishing whether Eq. \eqref{UniversalRel} is still valid for this entropy, the extreme value of the entropy is found looking for the value that causes
the temperature to vanish. This gives
\begin{equation}
    S_{R, ext} (\epsilon) = \frac{1}{\lambda} \ln \left( 1 + \lambda S_{0, ext} (\epsilon) \right),
\end{equation}
where
\begin{equation}
    S_{0,ext} (\epsilon) = \frac{\pi l^2}{6 \epsilon} \left\{ -1 + \sqrt{1 + \frac{12 Q^2 \epsilon}{l^2}} \right\}.
\end{equation}
Substituting this value into Eq. \eqref{Rrhs} we get the same expression obtained in the original reference, which is
\begin{equation}
    -T \left( \frac{\partial S_R}{\partial \epsilon} \right) \Big|_{M_{0,ext} (\epsilon)} = \frac{l}{12 \sqrt{6} \epsilon^{3/2}} 
    \left( -1 + \sqrt{1 + \frac{12 Q^2 \epsilon}{l^2}} \right)^{3/2}.
\end{equation}
This is also the case for the mass of the black hole, $M_{R, ext} = M_{0,ext} (\epsilon)$, explicitly given by
\begin{equation}
    M_{0, ext} (\epsilon) = \frac{12 Q^2 \epsilon + l^2 \left(-1 + \sqrt{1 + \frac{12 Q^2 \epsilon}{l^2}} \right)}{3 \sqrt{6 \epsilon} l 
    \sqrt{-1 + \sqrt{1 + \frac{12 Q^2 \epsilon}{l^2} }}}.
\end{equation}
Deriving this expression wrt $\epsilon$ we check that Eq. \eqref{UniversalRel} is satisfied.

We observe that this result is also true for the other non-extensive entropies considered here.





\end{document}